\documentclass[conference]{IEEEtran} 

\usepackage[T1]{fontenc} 
\usepackage{amsmath,amssymb,physics} 
\usepackage{algorithm}
\usepackage{algpseudocode}
\usepackage{float}
\usepackage{graphicx} 
\usepackage{cite} 
\usepackage{balance} 
\usepackage{qcircuit}
\usepackage{subcaption}
\usepackage{comment}
\usepackage{siunitx}
\usepackage{booktabs}
\usepackage{blindtext}
\usepackage{svg}
\usepackage{url}
\usepackage{xurl}

\usepackage{array}
\usepackage{balance} 
\usepackage{blindtext}
\usepackage{bm}
\usepackage{booktabs}
\usepackage{comment}
\usepackage{datetime}
\usepackage{fancyhdr}
\usepackage{float}
\usepackage{graphicx} 
\RequirePackage[colorlinks=true, allcolors=blue]{hyperref}
\usepackage{makecell}
\usepackage{multirow}
\usepackage{mwe}
\usepackage{siunitx}
\usepackage{subcaption}
\usepackage{svg}
\usepackage{tablefootnote}
\usepackage[absolute]{textpos}
\usepackage{url}
\usepackage{xcolor}
\usepackage{xurl}
\usepackage{enumitem}

\definecolor{ionqorange}{HTML}{FF5000}
\hypersetup{
    colorlinks=true,
    linkcolor=ionqorange,    
    citecolor=ionqorange,    
    urlcolor=ionqorange      
}

\IEEEoverridecommandlockouts 

\begin{document}

\title{Towards Scaling Quantum Fine-Tuning of Foundational Time Series Models for Classification}
\author{
    \IEEEauthorblockN{
        Sang Hyub Kim\IEEEauthorrefmark{1}\IEEEauthorrefmark{2},
        Julien Baglio\IEEEauthorrefmark{1}\IEEEauthorrefmark{3}\IEEEauthorrefmark{4},
        Rajiv Krishnakumar\IEEEauthorrefmark{1}\IEEEauthorrefmark{3}\IEEEauthorrefmark{4},
        Chi Chen \IEEEauthorrefmark{2},
        Oliver Knitter\IEEEauthorrefmark{2},
        Jonathan Mei\IEEEauthorrefmark{2},\\
        Claudio Girotto\IEEEauthorrefmark{2},
        Masako Yamada\IEEEauthorrefmark{2},
        Frederik F. Fl\"{o}ther\IEEEauthorrefmark{3}\IEEEauthorrefmark{4},
        Martin Roetteler\IEEEauthorrefmark{2}
        \medskip
    }
    \IEEEauthorblockA{
        \IEEEauthorrefmark{2}IonQ Inc., 4505 Campus Dr, College Park, MD 20740, USA\\ \{oliver.knitter, sang, jmei, claudio.girotto, \\ chi.chen, yamada, martin.roetteler\}@ionq.co}
    \IEEEauthorblockA{    \IEEEauthorrefmark{3}QuantumBasel, Schorenweg 44b, 4144 Arlesheim, Switzerland\\ \{julien.baglio, rajiv.krishnakumar, frederik.floether\}@quantumbasel.com}
    \IEEEauthorblockA{
        \IEEEauthorrefmark{4}Center for Quantum Computing and Quantum Coherence (QC2),
Department of Physics, University of Basel,\\ Klingelbergstrasse 82, 4056 Basel, Switzerland}
}

\maketitle

\begingroup
\renewcommand\thefootnote{\IEEEauthorrefmark{1}}

\makeatletter\def\Hy@Warning#1{}\makeatother
\footnotetext{Equal contribution}
\endgroup

\begin{abstract}
Time-series foundation models produce rich embeddings, but whether quantum models can exploit them, and how far hybrid classical-quantum architectures scale, remains unclear. We address this by fine-tuning Chronos for power-grid event classification (PSML-5) with a quantum head on the model's embeddings. Grouping embeddings by physical sensor type before summarization already surpasses the best published baseline built for this benchmark, and with finer-grained features the quantum head outperforms a larger classical multilayer perceptron on identical inputs by 1.7--2.0 percentage points of balanced accuracy. Yet the gains saturate: past a point, feeding more information to the same fixed-width register yields no improvement. We show the bottleneck is neither the supply of information nor circuit expressiveness, but the bandwidth of the data intake. To overcome this limitation, we introduce the \emph{wing} module, a self-contained few-qubit circuit that feeds additional information into the core circuit through a sparse, one-way coupling. Under a preregistered four-seed protocol, we attach wings to a fixed 12-qubit core with fixed features. Balanced accuracy increases with each added wing, from 83.6\% with no wings (13 qubits, including a post-selection qubit) to 85.2\% with two (19 qubits). Ablations establish that a circuit enlarged without new information gains nothing, while a wing fed information from the wrong sample harms accuracy. These results reframe scaling for quantum fine-tuning: added qubits help when they carry added inputs, not merely more parameters. Wings offer a modular and stable route to widening that bandwidth.

\end{abstract}

\section{Introduction}
Foundation models effectively amortize their large-scale pretraining costs by enabling straightforward adaptation to downstream tasks, using far less
data and compute than training from scratch~\cite{brown_language_2020,devlin_bert_2019}.
The standard method for adapting these models is through fine-tuning~\cite{jung_joint_2015,kading_fine-tuning_2017,wang_growing_2017}:
a small task-specific head is attached to the foundation model, whose pretrained weights are either frozen or lightly updated, and the resulting model is trained on a specialized
dataset for the new task. This is the
standard setup for large-language models (LLMs), such that the full capability of these  models can be efficiently applied to customized datasets without retraining the entire LLM from scratch. In the realm of time series forecasting, a similar type of prediction problem, foundation models such as Chronos~\cite{ansari2024chronos} and
TimesFM~\cite{das2024timesfm} tokenize, or otherwise embed, temporal
sequences and are similarly amenable to fine-tuning. Once pretrained on large forecasting datasets, they expose
hidden-state representations that are reusable for other tasks, not just those involving
unseen data, but also tasks these models were not originally trained to
perform~\cite{li2025trace}.

Quantum machine learning (QML), which replaces classical neural networks with parameterized quantum circuits (PQCs) that exploit fundamental properties of quantum mechanics, can offer new avenues to enhance fine-tuning capabilities beyond a potential asymptotic runtime advantage, specifically improvements in accuracy, parameter efficiency, and the ability to process
high-dimensional or otherwise difficult
data~\cite{peral2024systematic,peters2021machine,marshall2023high,aaronson2022much}. Because PQCs correspond to kernel machines with feature maps determined by the choice of encoding and ansatz~\cite{schuld_supervised_2021,perez-salinas_data_2020}, a natural place to insert them into a modern AI stack is as a fine-tuning head on top of a frozen foundation model. This setup also minimizes the amount of quantum trainable parameters needed to perform the desired task.

Prior work has explored a hybrid quantum--classical architecture that classifies language model embedding vectors using quantum-inspired encoders and a data-reuploading PQC classifier. A demonstrable accuracy advantage over similarly-sized classical baselines has been shown for binary sentiment classification~\cite{kim2025quantum}, and subsequent experiments successfully executed these circuits on a trapped-ion processor, with comparable classification accuracy in the presence of hardware noise~\cite{knitter2026energy}. This architecture distinguishes itself from parallel lines of research exploring the capacity to improve Low-Rank Adaptation (LoRA) fine-tuning using quantum circuits~\cite{koike2025quantum, kong2025quantum, liu2025quantum}, which are quantum-inspired methods explicitly simulating quantum circuits using state vector simulation or tensor networks~\cite{stoudenmire_supervised_2016}; like the quantum-inspired encoder in the hybrid model, these methods primarily perform dimension reduction and utilize the entire modeled state vector, so they cannot be efficiently executed using actual quantum hardware. In contrast, the hybrid fine-tuning approach in Kim et al.~\cite{kim2025quantum} explicitly utilizes quantum hardware with the goal of operating at scales that require exponential time and energy cost to simulate~\cite{knitter2026energy}.

A natural question, left open so far, then is to ask whether this approach generalizes beyond text, and more importantly, how it should scale.
To that end, this paper seeks to determine whether quantum
fine-tuning can be effectively applied to a time-series foundation model in order to solve a classification task, alongside how to appropriately scale the model in this regime: can a PQC improve over a classical head? Does the quantum performance reliably increase as more qubits are utilized while the circuit is simultaneously kept shallow enough for noisy intermediate-scale quantum (NISQ) hardware?
Time-series classification differs from the previously tested setting in several respects that can stress the classification head:
\begin{enumerate}[label=(\roman*)]
    \item The inputs are multivariate, often with tens of sensor channels and hundreds of time steps.
\item The downstream task is frequently multi-class, rather than just binary.
\item The relevant structure may live in short transients superimposed on large nuisance variation.
\end{enumerate} 

To make our experiments representative of these three difficulties, we choose the PSML power system benchmark~\cite{zheng2022psml}, a multivariate time series
classification benchmark drawn from power grid monitoring as would be relevant to the energy sector. It contains physically distinct disturbance types, such as generator
tripping, branch tripping, bus tripping, branch faults, and bus faults. The input data consist of millisecond-level synchrophasor measurements, and the disturbance types must be recovered from these measurements despite the same event class being realized at electrically distinct locations across a topologically diverse grid.

We therefore organize our classification study around five of the PSML classes: branch fault, branch trip, bus fault, bus trip, and gen trip. This task is referred to as PSML-5 in the rest of the paper. We attach classification heads to a frozen Chronos-T5
backbone, implementing both a quantum head and a larger multilayer perceptron
(MLP) classical variant. Directly comparing the quantum head with the MLP, we investigate
how its classification accuracy scales with the number of qubits. The classical head also lets us benchmark the
forecasting-to-classification fine-tuning setup against some of the specialized
architectures reported in the original PSML study (for example convolutional neural networks such as InceptionTime~\cite{Fawaz2020}, or long short-term memory models such as
MLSTM-FCN~\cite{karim2019}), establishing that this conversion can
beat the best time series models trained for the same task.

We can also test whether additional classical features, in
the form of \emph{sketch fingerprints} representing compact summaries of the
multivariate window, can improve a quantum head of fixed width, or
if the head saturates once its register is full. If the latter
holds, the bottleneck must be the head's intake capacity, not the availability of input information. Growing
intake capacity naively by deepening the circuit or widening a fully-connected
register would increase two-qubit gate count and place additional pressure on trainability. To address this issue, we introduce a new architecture where we attach \emph{wings}: three-qubit blocks that receive their own per-channel feature streams, supplementing a core that keeps its compact fingerprints fixed. Each wing influences the core circuit only through sparse, one-way entangling gates. The depth of the core is deliberately held constant.
Wings are the mechanism by which qubit count is scaled in this work, enabling an increase in circuit intake without increasing depth or fully-connected width.
Our core PQC is a 12-qubit head; together with a dedicated post-selection qubit it forms the 13-qubit, zero-wing rung, and is scaled
up to 19 qubits via two wings. 
Using bit-identical core features across variants allows us to attribute potential gains via ablations that grow circuit capacity without new information, and that shuffle wing payloads across samples.

The main contributions of this work are as follows.
\begin{itemize}
    \item We extend quantum fine-tuning to time series foundation models by
      converting a forecasting model into a multi-class event
      classifier on an energy dataset (PSML). We demonstrate that this setup surpasses the best specialized classical baseline
      originally reported for that dataset.
    \item We study the scaling of a fixed-width quantum head as
      additional sketch fingerprints are provided, testing whether the quantum head can
      outperform a larger classical MLP on average, and whether intake capacity is the main limiting factor, rather than feature supply.
    \item We introduce the wing, a novel architecture that grows intake at
      constant core depth and operates as the primary scaling mechanism (in qubit count) for our architecture. We compare models with different wing counts, reporting a qubit scaling ladder together with ablations that isolate payload information
      from added circuit capacity.
\end{itemize}

The remainder of this paper is organized as follows.
Section~\ref{sec:methodology} introduces the PSML dataset and its
classification problem, describes the baseline benchmark, presents the
Chronos backbone, and specifies the training pipeline and wing architecture. 
Section~\ref{sec:results} reports hyperparameter selection,
classification results, and the qubit-scaling ladder.
Section~\ref{sec:conclusion} presents the implications of
the results and a short conclusion.

\section{Methodology}
\label{sec:methodology}

\subsection{Dataset and Baseline}
\label{sec:methodology:dataset}
We evaluate on PSML, a multivariate time-series classification benchmark drawn from
power-system monitoring, which comprises synthetic but physically consistent
millisecond-resolution phasor measurement unit (PMU) recordings from a coupled
transmission--distribution grid~\cite{zheng2022psml}. Captured across these multi-scale
sensor signals are key dynamic disturbance events, such as generator trips and bus
faults, which automated control systems must accurately classify in real time to isolate
physical failures, prevent cascading blackouts, and maintain grid stability under
increasing renewable energy integration. To ensure a fair comparison to previous
benchmarks, we follow the same data pre-processing protocol as Zheng et
al.~\cite{zheng2022psml}. We use the five-class event-classification benchmark subset
aligned with the TAMU reference data loader~\cite{tamu_psml_dataset}. This subset comprises
Natural Oscillation scenarios only, excluding forced oscillation cases, and skips row
328 to match the reference processing. Each sample is a multivariate window of shape
$(91,960)$, where the 91 channels correspond to PMU-type quantities (e.g., bus voltage
magnitudes/angles and branch active/reactive power) over a 960-step millisecond event
window. Labels are five discrete disturbance types, mapped to integers $0-4$ as in the
TAMU benchmark: generator trip, branch trip, branch fault, bus trip, and bus fault, where
a \textit{trip} is the intentional opening of a circuit breaker to isolate a component (a
generator, transmission line/branch, or station bus), whereas a \textit{fault} is an active
short circuit or physical disruption (like a lightning strike, fallen tree, or insulation
breakdown) occurring directly on a branch or bus. After filtering, the benchmark contains
549 samples. Following the reference protocol, we use a deterministic 80/20 split by sorted
row index (first $80\%$ train, last $20\%$ test), yielding 439 training and 110 test
samples: this is hence not a random stratified split. Approximate class counts are:
\begin{enumerate}
\item \textit{Train set:} generator trip 141, branch trip 82, branch fault 75, bus trip 72, bus fault 69;
\item \textit{Test set:} 43, 19, 17, 14, 17 respectively.
\end{enumerate}

As a baseline, we compare against the top result reported by Zheng et al.~\cite{zheng2022psml}, obtained using the MLSTM-FCN architecture~\cite{karim2019}. They have evaluated performance using balanced accuracy, defined by
$(\mathrm{sensitivity}+\mathrm{specificity})/2$, rather than standard accuracy to account
for severe class imbalance, where routine operating conditions far outnumber rare dynamic
disturbance events. Across ten random initializations of the model weights on the
designated dataset split, this baseline achieved a mean score of $74.2\% \pm 2.9\%$.

We select PSML specifically for the structural difficulty it presents: a given disturbance type is simulated at many electrically distinct locations across a topologically diverse grid, so instances of the same class differ substantially in baseline level, onset timing, and peak amplitude, while an average class-level signature nonetheless remains detectable.
The central challenge is therefore robustness to substantial nuisance variation around a
still-recoverable signal, rather than the absence of a separating signal altogether. A
supporting qualitative and quantitative analysis, including example waveforms, a
low-dimensional projection, and a class-separability measure, is provided in
Appendix~\ref{app:dataset-structure}.

\subsection{Foundation Models}
\label{sec:methdology:models}
We use Chronos~\cite{ansari2024chronos} as the pretrained time-series foundation model
underlying all classification heads reported in this paper. Chronos tokenizes scaled,
quantized time-series values into a fixed vocabulary and is trained on a large corpus of
real and synthetically generated series to perform probabilistic forecasting. We use the
Chronos-T5 variants, which adopt the encoder--decoder T5 architecture and are released in a range of sizes from tiny (8M parameters) to large (710M parameters); they expose
per-timestep hidden-state representations that we take from the encoder stack and use as
frozen features for a downstream classification head (Section~\ref{sec:methdology:heads}),
rather than fine-tuning the foundation model's own weights. We use the base variant (200M
parameters, $d_\mathrm{model}=768$) throughout, extracting embeddings independently per
channel and concatenating across channels for the multivariate dataset described in
Section~\ref{sec:methodology:dataset}.

Chronos was selected over TimesFM~\cite{das2024timesfm}, a comparably sized (200M-parameter)
\emph{decoder-only} forecasting foundation model from Google that we evaluated under the same
feature-extraction and fine-tuning protocol, on the basis of a modest but consistent edge in
downstream classification performance across our preliminary experiments. We emphasize that
this comparison was not exhaustive.

In particular, we did not evaluate Chronos-2~\cite{ansari2025chronos2}, a more recent release from the same family that adds native multivariate and covariate support, nor other candidate time-series foundation models (e.g., Lag-Llama~\cite{rasul2024lagllama}, TimeGPT~\cite{garza2024timegpt}). Since the quantum fine-tuning architecture is agnostic to the foundation model, we consider a systematic comparison across backbones to be outside the scope of the present work, though it remains an important direction for future work to establish how sensitive the reported gains are to the choice of underlying foundation model.

\begin{figure*}[hbt]
    \centering\includegraphics[width=1\linewidth]{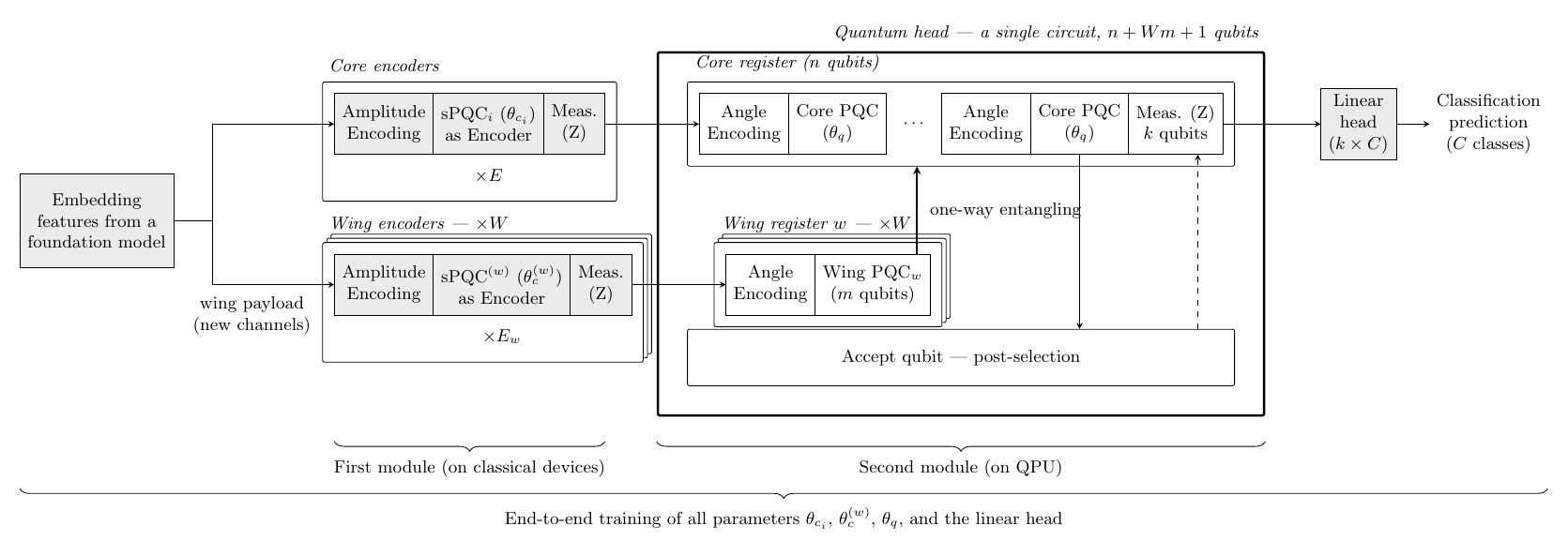}
    \caption{The wing architecture. Classical encoders (shaded; sPQC = simulated
    parameterized quantum circuit) are separate per module, while the second
    module is a single quantum circuit whose register is partitioned into a
    core, the wing registers, and one accept qubit. The core ($n$ qubits, $E$
    encoders, $R$ data re-uploading blocks, fixed depth) carries all readout:
    Pauli-Z on $k$ qubits feeds a $k\times C$ linear head. Each wing (one
    register and $E_w$ encoders per wing, $\times W$) adds $m$ qubits and
    couples to the core only through a few one-way entangling gates (fixed CNOTs in this work) at
    staggered points; wings never couple to each other. The accept qubit,
    coupled late to the core, is measured and only $|0\rangle$ outcomes are
    kept (post-selection). Total $n+Wm+1$ qubits. In this work $E{=}6$,
    $E_w{=}3$, $n{=}12$, $m{=}3$, $R{=}2$, $k{=}5$, $C{=}5$, and
    $W\in\{0,1,2\}$ gives $13, 16, 19$ qubits. Shaded blocks are classical;
    the quantum circuit is executed in noiseless simulation in this work
    (the unshaded module is the part intended for a QPU).}
    \label{fig:full_block_diagram}
\end{figure*}

\section{Training \& Fine-Tuning Pipeline}
\label{sec:methdology:heads}

Once all of the 91 PMU-type channels of a PSML-5 window are tokenized and passed independently through the frozen Chronos-base encoder presented in Section~\ref{sec:methdology:models}, the resulting per-timestep hidden states are mean-pooled over the 960-step window, giving one 768-dimensional embedding per channel. These per-channel embeddings are then concatenated into a single raw feature vector. These steps are represented by the leftmost block in \autoref{fig:full_block_diagram}. From here, the classical and quantum architectures diverge.

On top of this shared representation we build in parallel a classical head, consisting of a multilayer perceptron (MLP) whose swept configurations carry 1.2--3.0 million trainable parameters, versus 3{,}240 for the 12-qubit quantum head  (the sPQC encoders, the circuit's rotation weights, and the linear readout combined), and a quantum head implementing the architecture from~\cite{kim2025quantum} that is central to this work. While the classical MLP consumes the (pooled) feature vector directly, the quantum head requires the feature vector to be compressed into a suitable width for a NISQ qubit register. We evaluated two encoder choices for this compression step: a classical (dense) encoder, and a simulated-PQC (sPQC) encoder that models an amplitude encoding into a small parameterized circuit, read out via Pauli-Z measurement. The sPQC encoder consistently outperformed the classical one, which is why \autoref{fig:full_block_diagram} shows only the sPQC variant. The sPQC encoder is used identically for the core encoders and for each wing encoder.

Having fixed the core quantum head architecture, we consider whether additional, manually engineered features could improve either head. We derive these from a type-group mean/std pooling of the Chronos embedding channels, grouped by PMU measurement type (voltage, active power, reactive power) and reduced to a per-group mean and standard deviation, augmented with sketch fingerprints, which are compact, deterministic per-block summaries designed to recover local or transient structures that mean/std pooling alone smooths out. More details of these features can be found in Appendix~\ref{app:manual-features} (Eqs.~(\ref{eq:rms_std})--(\ref{eq:energy_split})). For the classical head, incorporation is simplest: the fingerprints are combined with the pooled feature vector into one flat input, with no architectural distinction between base and engineered features, and fed to the MLP as-is. For the quantum head, we first test the same sketch fingerprints as part of the core's own input, growing the classical vector each core encoder consumes, without changing qubit count, to test whether a fixed-width quantum head saturates prematurely.

Afterward, we scale qubit count directly via wing extensions, shown in \autoref{fig:full_block_diagram}. For these experiments, the core retains the 96-fingerprint feature set selected in Section~\ref{sec:classification_results} (the base sketch augmented with the raw-channel sketch; Appendix~\ref{app:sketch_fingerprints}), held bit-identical across all rungs; each wing then carries its own separate feature
stream. Each wing is a register with its own sPQC encoder that receives a feature stream built from the ungrouped per-channel embedding features within one PMU-type group (before the type-group pooling), later replaced with a standardized sketch of these same raw channels, which are constructed analogously to the core sketch fingerprints (see Appendix~\ref{app:sketch_fingerprints}) but applied to the per-channel embedding stream rather than the type-group mean/std. Throughout, we refer to fingerprints computed from the pooled blocks as the \emph{base sketch}, and to those computed from the per-channel streams as the \emph{raw-channel sketch}; feature sets that include both are the \emph{base+raw} family. Each wing couples into the fixed-depth core through a small number of one-way entangling gates, implemented as fixed CNOT gates carrying no trainable parameters, so what a wing contributes is determined entirely by its own trained encoder and payload. These couplings are placed at staggered points in the circuit: wings couple successively, interspersed with mixing layers within the core. Wings never couple to each other, and core depth is held fixed as wings are added. Scaling qubit count via wings therefore adds intake capacity and payload without adding depth or widening a fully-connected register.

A dedicated post-selection qubit is present at every rung of the qubit-scaling ladder so that readout grammar is held fixed. It is coupled late into the core through trainable controlled-$R_Y$
rotations deliberately initialized near zero, at $\theta_0=0.05$, so the acceptance mechanism opens gradually over
training. It is then measured, and only the $|0\rangle$ outcome is kept (post-selection). Because a low acceptance rate wastes samples and increases variance, at every rung that includes the post-selection qubit, the training loss adds a per-sample penalty that discourages acceptance rates below a fixed floor of 0.25. Data re-uploading is applied with $R=2$ repetitions throughout, while circuit depth is fixed per module and does not grow as wings are added, and readout is a local $Z$ measurement on five core wires feeding a linear head. With the 12-qubit core plus the post-selection qubit, the ladder starts at 13 qubits. All circuits are simulated noiselessly.

\section{Results}
\label{sec:results}

\subsection{Hyperparameter optimization}
\label{sec:results:hyperparameters}

The initial hyperparameter search used Optuna's Tree-structured Parzen Estimator (TPE) search \cite{akiba2019optuna}, using five-fold cross-validation on the training split, with mean balanced accuracy as the selection objective; similarly, the values reported in \autoref{tab:psml5_rawish_sketch_appendix} are five-fold cross-validation means on the training split. Throughout, folds are defined using scikit-learn \texttt{KFold(n\_splits=5, shuffle=True, random\_state=seed)} on the 439-sample training split. A reported value for a single run is the mean balanced accuracy over its five validation folds (its fold-mean), quoted where shown with the standard deviation across folds. Changing the seed re-draws both the fold assignment and the network initialization, so repeated seeds constitute repeated five-fold cross-validation; ladder entries aggregate four seeds (42--45) as the mean of the four fold-means, quoted with the standard deviation across seeds. For the classical head, the search space covers learning rate, dropout, weight decay, batch size, learning-rate schedule, and eight candidate hidden-layer architectures (see Appendix~\ref{app:hparam_classical}). For the quantum head, core only, the search space additionally covers the classical pre-encoder width, qubit count, circuit depth (number of data-reuploading layers), number of measured qubits, entangling-gate connectivity, and post-selection setting (see Appendix~\ref{app:hparam_quantum_core}), and was further refined across several manual rounds following the initial automated search (see Appendix~\ref{app:hparam_refinement}).

A first extension, sketch fingerprints on the quantum head, reuses the unchanged 12-qubit core configuration, varying only the sketch-fingerprint dimension as the reported ablation axis. We performed a controlled ablation rather than an independent hyperparameter search. The classical head with engineered features, by contrast, is tuned exactly like the base classical head: the same search space (see Appendix~\ref{app:hparam_classical}) is searched again for the longer, sketch-augmented input, with all parameters of the network updated jointly by a single optimizer and no partial freezing.

\subsection{Classification Results}
\label{sec:classification_results}

\begin{figure}[t]
\centering
\includegraphics[width=\linewidth]{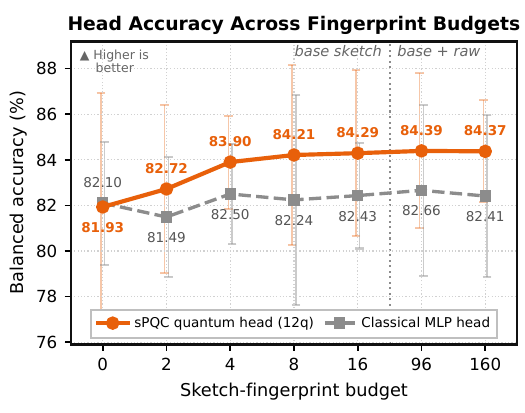}
\caption{Balanced accuracy of the sPQC quantum head and the classical MLP head vs.\ sketch-fingerprint budget (best configuration per budget; error bars: fold standard deviation). The dotted divider marks the family switch: budgets up to 16 use the base sketch only, 96 and 160 use the base+raw family. Values are five-fold cross-validation means
(\autoref{tab:psml5_rawish_sketch_appendix}).}
\label{fig:sketch_ablation}
\end{figure}

We show in \autoref{fig:sketch_ablation} the balanced accuracy we obtain with both the classical and quantum heads, as a function of the sketch-fingerprint budget, all under the five-fold cross-validation protocol of Section~\ref{sec:results:hyperparameters}. The corresponding numbers are found in \autoref{tab:psml5_rawish_sketch_appendix} (Appendix~\ref{sec:psml5_rawish_appendix}) which reports balanced accuracy across the full progression from the published benchmark to the sketch-fingerprint
ablation, with each entry being the best
configuration found per fingerprint budget. Even before any of the
feature engineering steps described in Section~\ref{sec:methdology:heads},
replacing the specialized PSML architectures with a frozen Chronos-base
backbone and a plain classical head (E0) already matches the published
benchmark (74.43\% vs.\ 74.20\%, B0). As preregistered, a one-shot evaluation on the untouched 110-sample test
split is reported in Appendix~\ref{app:test_eval}; there, the ensemble
confidence intervals of the 13- and 19-qubit rungs also lie entirely
above the published test baseline. Moving to the type-group mean/std
pooling of E1 (Appendix~\ref{app:type_group_pooling}) produces the
largest single jump in the table for both heads, to roughly 82\%
balanced accuracy (81.93\% quantum, 82.10\% classical); at this
zero-fingerprint budget the two heads are comparable.

From a fingerprint budget of two onward (E2; this budget mixes connectivity settings, whereas E3--E7 are uniform sweeps), the quantum head leads the
classical head on bit-identical cached inputs. Both heads
saturate as budget grows further, but at different levels: the quantum
head's saturation trajectory (84.21\% $\to$ 84.29\% $\to$ 84.39\% $\to$
84.37\% at budgets 8, 16, 96, and 160 respectively) settles into a
plateau roughly 1.7--2.0 percentage points above the classical head's
own, lower plateau, meaning the same information converts into less accuracy
for the classical head than for the quantum one. This plateau is not
driven by a single continuously growing feature family: budgets 2--16 use the base sketch only, while budgets 96 and 160 switch to the base+raw family, which additionally incorporates the raw-channel sketch (\autoref{sec:methdology:heads}). The 16$\to$96 point
is therefore a family switch, not simply, ``more of the same
fingerprint,'' and the near-flat accuracy across this switch (84.29\%
at 16 vs.\ 84.39\% at 96) suggests the quantum head's fixed 12-qubit
width, rather than the specific feature family being used, is the binding
constraint at this point in the ablation. The best-performing fixed-width configuration (96 sketch fingerprints,
84.39\% $\pm$ 3.40\% balanced accuracy) is the 12-qubit core
configuration used as the fixed starting point for the
qubit-scaling experiments in Section~\ref{sec:qubit_scaling}.

\subsection{Qubit Scaling}
\label{sec:qubit_scaling}

The 12-qubit core saturates as more input features are loaded into its
fixed register, as seen in Section~\ref{sec:classification_results}, motivating the wing architecture of Section~\ref{sec:methdology:heads}. To isolate the effect of qubit count, we hold the readout grammar fixed and vary only the number of attached three-qubit wings---zero, one, and two wings---with the plain 12-qubit core kept as an external
reference point using a different apparatus, without post-selection. Note that with zero wings we have a 13-qubit PQC as we always include the post-selection qubit in the scaling experiment. 

Each rung is evaluated at its own best operating point under an
identical search protocol: a bracket search on seed 42 selects the peak
learning rate, and the peak together with its two neighboring values is
then evaluated across seeds 42--45, each re-drawing both the five-fold
cross-validation split and the network initialization, under identical
training budget, data splits, and selection criteria; the training protocol is given in Appendix~\ref{app:hparam_wings} and the reported learning-rate windows in Appendix~\ref{app:qubit_ladder_full}. All reported values are
internal validation balanced accuracy under five-fold cross-validation; no
test data is used at this stage.

\begin{figure}[t]
\centering
\includegraphics[width=\linewidth]{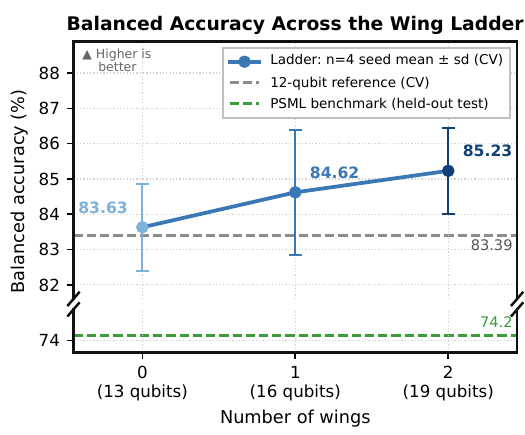}
\caption{Balanced accuracy vs.\ number of wings. Points: mean over seeds
42--45 at each rung's peak learning rate (selected on the seed-42 sweep);
error bars: standard deviation across the four seed fold-means. Dashed
line: the 12-qubit reference (different apparatus, bit-identical core features). Ladder and reference are internal five-fold cross-validation
on the training split. The dotted green line is the published PSML-5
baseline (best reported result) of $74.2\%$~\cite{zheng2022psml}, the mean over ten random
initializations of the model weights on the designated train/test split.}
\label{fig:qubit_ladder}
\end{figure}

We present in \autoref{fig:qubit_ladder} the qubit-scaling (wings scaling) of the balanced accuracy (in solid blue), reporting the standard deviation across the four seed fold-means. The gray dashed line is the 12-qubit core PQC reference to visualize the starting point of our wing mechanism. The dotted green line at $74.2\%$ is the published MLSTM-FCN baseline, the best reported result on PSML-5~\cite{zheng2022psml}. That mean is taken over ten random initializations of the model weights on Zheng et al.'s designated train/test split; the published uncertainty ($\pm 2.9\%$, not drawn here) is therefore not comparable to the ladder error bars. We also report in \autoref{tab:qubit_ladder_full} (Appendix~\ref{app:qubit_ladder_full}), for each rung, the mean balanced accuracy across four seeds at the peak learning rate (the
preregistered primary reading) together with the per-seed values, whose maxima serve as an auxiliary best-run reference. Balanced accuracy increases monotonically with wing count: 83.63\% (13 qubits, zero wings) $<$ 84.62\% (16 qubits, one wing) $<$ 85.23\% (19 qubits, two wings). The 12-qubit reference sits below all three ladder rungs at 83.39\%. The starred peak is selected on the seed-42 curve; the $n{=}4$ mean is marginally higher at learning rate 0.010 ($83.53\%$). This 12-qubit reference is evaluated under the same preregistered scan protocol as the ladder rungs, and is distinct from the 96-fingerprint 12-qubit configuration of Section~\ref{sec:classification_results} (84.4\%), which was selected under a different, wider search.

As a check that this ordering is not an artifact of per-rung
operating-point selection, we also report a single learning rate
(0.008) shared across all four configurations, with no per-rung tuning:
12-qubit reference 83.01\% $<$ 13q 83.63\% $<$ 16q 84.08\% $<$ 19q
84.87\%. Because 0.008 is close to optimal for the smallest rung and
off-peak for the larger ones, this comparison is biased against the
ordering it nonetheless preserves; we label it in
\autoref{tab:qubit_ladder_full} as a selection-free comparison.

A small number of caveats apply to this scan. Two outer points of the 13-qubit seed-42 bracket sweep (learning rates 0.006 and 0.010, outside the reported peak$\pm$1 window) carried a truncation flag in our training-length audit, with improvement still arriving late relative to the early-stopping patience; the 0.010 point also showed an elevated post-selection acceptance rate (39.1\% vs.\ 23.5--24.6\% at the reported learning rates). The 12-qubit reference's own seed-42 curve is non-unimodal, with a local dip at learning rate 0.008 (its lowest tabulated seed-42 cell). In the auxiliary best-run reading (the maximum per-seed value, not the primary $n{=}4$-mean), the 12-qubit reference's single best run can exceed the zero-wing, 13-qubit rung's best run. This is both expected and consistent with our central claim: the wings, not the fixed readout apparatus, are the source
of the observed gain.

\subsubsection{Attribution Ablations}
\label{sec:attribution_ablations}

Having established that balanced accuracy increases with wing count, we
next search for an explanation: does the two-wing circuit simply possess more trainable capacity, or is it specifically the information carried in the wing payload that makes a difference? We isolate this with two ablations at the two-wing configuration, each changing exactly one factor of the model while holding every other input, protocol, and training budget fixed, and each variant is evaluated at its own best learning rate over multiple seeds.

The first ablation grows circuit capacity without adding any new
information: the model's data-reuploading block is un-shared across its
repetitions, increasing the circuit's rotation-parameter count from 150 to 204 (one component of the two-wing model's total parameter count; the sPQC encoders and readout are untouched), with weights initialized so the un-shared circuit is
identical to the shared one at the start of training, while
input features remain identical to the intact two-wing model. If the
wing's gain were simply a matter of circuit size, this variant should
improve on the baseline, but if the gain instead comes from the injected
information, then it should not. We find no gain from this variant. Balanced accuracy drops
from 85.23\% (baseline) to 83.85\% ($\Delta = -1.38$ points), and this
is not a borderline result: the seed-42 learning-rate curve is
unimodal with a peak at 0.008 (83.66\% $\to$ 83.92\% $\to$ 84.57\% $\to$
84.41\% at learning rates 0.006--0.009), and the deficit relative to
baseline is consistently negative ($-1.4$ to $-1.9$ points) at every
candidate learning rate.

The second ablation instead breaks the correspondence between each
sample and its wing payload. Wing input channels are shuffled across
samples, meaning every wing still receives an input of identical
dimensionality, distribution, and parameter count, but the information
no longer matches the sample it is paired with. Core inputs are left
intact. If the wing's gain comes from the payload's information content
rather than its mere presence, this should remove the gain and the
model should behave like the zero-wing configuration, which is precisely what we observe. Balanced accuracy falls to 81.59\%, a fold-paired deficit of $4.20 \pm 3.92$ points relative to the intact two-wing model, computed over the 15 folds shared by both variants (three common seeds). Thirteen of the 15 paired folds are negative, one is a tie, and one is positive by 0.47 points. A paired $t=-4.15$ ($n{=}15$, $p\approx0.001$) is reported as a descriptive statistic, since folds within a seed share training data, and the fully independent seed-level comparison is unanimously negative across all three common seeds ($-5.22$, $-4.56$, $-2.82$ points). More strikingly, the shuffled-payload model (81.59\%) performs below the zero-wing, 13-qubit configuration itself (83.63\%, a further $-2.04$ points), indicating a wing carrying broken information fails to match the performance of a wingless model.

Taken together, these two ablations indicate that the benefit of the
wing is carried by the information in its payload, and not by the added
circuit capacity. Destroying only the sample-correspondence of the
payload, while still leaving its distribution, circuit, and parameter count
unchanged, not only removes the gain but underperforms the wingless
model.

\section{Discussion and Conclusion}
\label{sec:conclusion}
This work answers two central questions: whether quantum fine-tuning successfully transfers to a time-series foundation model, as shown by beating both specialized time-series classification models and classical fine-tuning of a time-series foundation model, and how the resulting quantum head should scale. We have found that training a quantum head on the embeddings of a frozen Chronos backbone beats the strongest baseline built for PSML-5, and that once finer detail is added, the quantum head converts identical inputs into $1.7$--$2.0$ percentage points more balanced accuracy than a larger classical counterpart. In answering the scaling problem, we have found that a fixed-width intake saturates as it receives richer information: beyond a modest fingerprint budget, additional features no longer improve accuracy. In response, we grow the intake itself with a new wing architecture. Our experiments show that balanced accuracy rises monotonically with wing count under a preregistered protocol, and ablations attribute this gain to the information the wings carry; a circuit enlarged without new information gains nothing, while a wing fed information from the wrong sample performs worse than no wing at all. We have thus demonstrated a scaling mechanism where adding more qubits effectively leads to an increase in accuracy, without compromising the shallowness of our core 12-qubit PQC, keeping training effective.

Several limitations bound these claims. The attribution ablations were run at the two-wing (19-qubit) operating point only. All reported numbers are internal five-fold cross-validation results from state-vector simulation using a single benchmark. The trainability argument for the sparse one-way coupling is made at the design level, through constant per-module depth and restricted connectivity, and is supported by the ladder training successfully at every rung; we have not performed a quantitative gradient study. An observation we did not anticipate concerns initialization. In the capacity-growth ablation, the enlarged circuit begins training as an exact functional copy of its smaller baseline, the safest imaginable head start, yet it still ends below that baseline, as seen in Section~\ref{sec:qubit_scaling}. In our setting, we therefore do not rely on inherited solutions: every grown model in the ladder is trained from scratch. 

Natural next steps include executing the trained heads on quantum hardware and characterizing shot-noise sensitivity in the inference task, extending the ladder beyond two wings, replicating the ablations at other scales, a systematic study of initialization for grown models, evaluating further datasets, and a gradient-based trainability analysis of the wing-extended circuits. Altogether, we conclude that quantum fine-tuning does transfer to time-series foundation models, having demonstrated that it surpasses the strongest baseline built for PSML-5. On identical inputs, a quantum head also converts sketch fingerprints into higher balanced accuracy than a classical head on the same frozen backbone. To improve these results further, we introduced a wing mechanism that scales the quantum head not by enlarging the circuit, but by growing input bandwidth, showing that accuracy increases as wing/qubit count increases.

\section*{Acknowledgment}
Generative AI tools were used to assist with code implementation, debugging, experiment scripting, data analysis support, and drafting or editing portions of the manuscript text. All scientific claims, experimental design, result verification, and interpretation were carried out and approved by the authors.

\bibliographystyle{ieeetr}
\bibliography{bibliography}

\appendices

\section{Structural Characterization of PSML}
\label{app:dataset-structure}

This appendix expands on the structural difficulty of PSML referenced in
Section~\ref{sec:methodology:dataset}: PSML instances exhibit high intra-class variance
because a given disturbance type is simulated at many different, electrically distinct
locations across a topologically diverse grid. We support this characterization with example
waveforms, a low-dimensional projection, and a quantitative class-separability measure.

\begin{figure}[t]
\centering
\includegraphics[width=\linewidth]{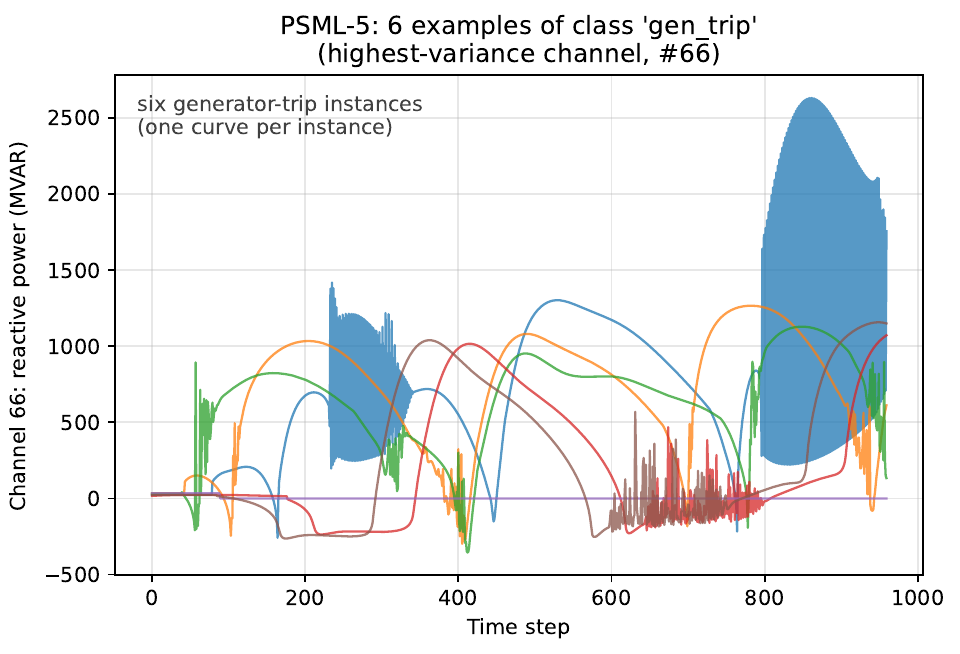}
\caption{Six PSML generator trip instances overlaid on the dataset's highest-variance channel (\#66, reactive power).
Despite the shared event label, they vary widely in baseline, onset timing, and amplitude.}
\label{fig:app-overlay}
\end{figure}

\autoref{fig:app-overlay} illustrates the intra-class variance directly: six generator
trip instances, plotted on their most active PMU channel, show markedly different baselines,
oscillation onset times, and peak amplitudes---ranging from one instance that remains
essentially flat on this channel to another exhibiting a sustained oscillation exceeding
2500 in the recorded units---despite sharing an identical event label. This is the expected
consequence of the same event type occurring at different points across the co-simulated
23-bus transmission system and two 13-bus distribution feeders, each with distinct local
impedance and downstream load structure.

\begin{figure}[t]
\centering
\includegraphics[width=\linewidth]{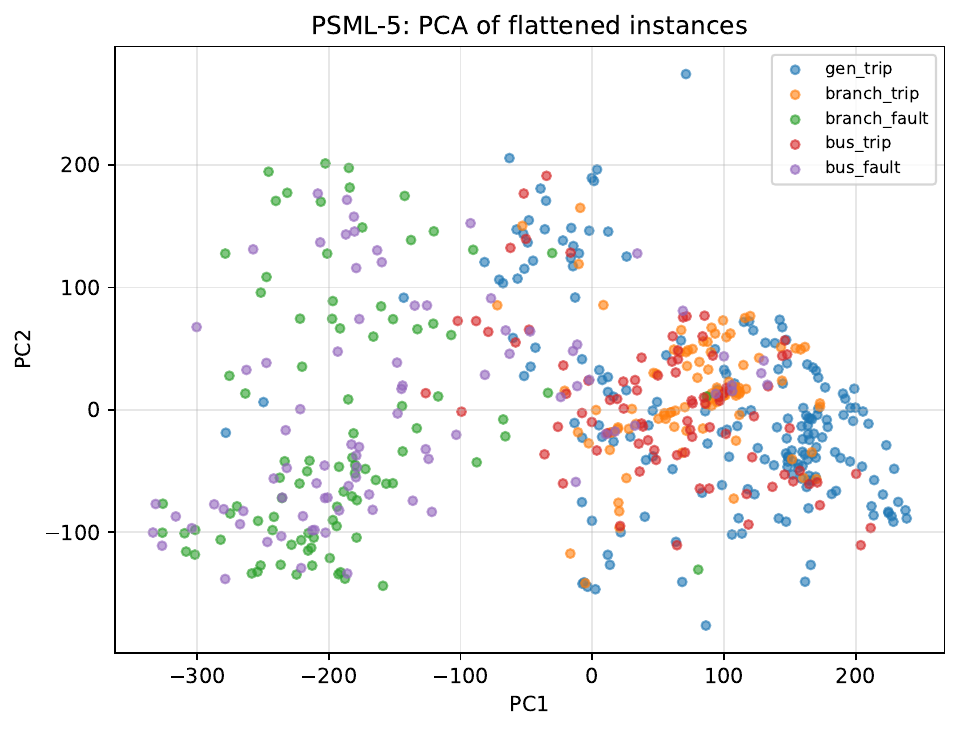}
\caption{PCA projection (2 components) of flattened, standardized PSML instances, colored by
class. PC1 predominantly separates fault-type from trip-type events; classes within each
group overlap substantially.}
\label{fig:app-pca}
\end{figure}

The two-dimensional PCA projection of flattened, standardized instances in \autoref{fig:app-pca} shows that this instance-level diversity does not destroy class
structure at the dataset level, but that the structure is coarser than the five-way labeling
suggests. The first principal component predominantly separates \emph{fault}-type from
\emph{trip}-type events: branch fault and bus fault instances occupy the negative-PC1 region
and overlap heavily with one another, while the three trip classes occupy positive PC1, with
branch trip and bus trip concentrated in a dense central band that also overlaps
substantially with generator trip. The leading linear directions therefore encode the
physical distinction between an active short circuit and an intentional breaker opening more
strongly than they encode \emph{which} component the event occurred on, consistent with a
task in which event location acts largely as nuisance variation.

To quantify class separability beyond visual inspection, we flattened and standardized the
instances, reduced them to 50 principal components, and computed all pairwise Euclidean
distances in this space. From the resulting distance matrix we derive the mean pairwise
distance between same-class instances (intra-class) and between different-class instances
(inter-class), each normalized by the dataset's overall mean pairwise distance so that the
values are scale-free and directly interpretable relative to the dataset's own average. We
repeat this measurement with an outlier-robust (median/IQR-based) scaler in place of standard
scaling, to check whether the pronounced amplitude spread visible in
\autoref{fig:app-overlay} and the small number of extreme instances visible in
\autoref{fig:app-pca} are driving the result. \autoref{tab:app-separability} reports
both versions.

Under both scalers the inter-to-intra-class distance ratio remains above 1 (1.13--1.17), and
intra-class distances sit below the dataset's overall average (0.88--0.91): despite the
pronounced instance-to-instance waveform diversity visible in \autoref{fig:app-overlay},
an average class-level signature remains detectable even under a simple linear projection,
and this conclusion survives down-weighting the extreme instances. High intra-class variance
and non-trivial class separability are therefore not mutually exclusive: PSML's principal
challenge is robustness to substantial nuisance variation (event location and local grid
topology) around a still-recoverable signal.

\begin{table}[t]
\centering
\caption{Intra-/inter-class pairwise-distance ratios for PSML-5 (normalized by the dataset's
overall mean pairwise distance) in a 50-component PCA projection of flattened, scaled
instances. Separability = inter-class / intra-class distance; values $>1$ indicate classes
are, on average, farther apart than same-class pairs.}
\label{tab:app-separability}
\begin{tabular}{lcc}
\hline
 & Standard scaling & Robust scaling \\
\hline
Intra-class / overall      & 0.883 & 0.908 \\
Inter-class / overall      & 1.033 & 1.026 \\
Separability (inter/intra) & 1.171 & 1.131 \\
\hline
\end{tabular}
\end{table}

\section{Feature representation: type-group pooling and sketch fingerprints}
\label{app:manual-features}

\subsection{Type-Group Mean/Std Pooling}
\label{app:type_group_pooling}
Each PSML-5 channel name carries a measurement-type prefix, and channels are partitioned by this prefix into three groups: VOLT (bus voltage magnitude/angle), POWR (branch active power), and VARS (branch reactive power, "volt-amperes reactive"). Let $e_c \in \mathbb{R}^{768}$ denote the mean-pooled Chronos-base embedding of channel $c$ (Section II-C), and let $I_g$ be the index set of channels belonging to group $g \in \{\text{VOLT}, \text{POWR}, \text{VARS}\}$. We replace the full per-channel concatenation ($91 \times 768 = 69{,}888$ dimensions) with six group-level summary vectors,

\begin{equation}
    \mu_g = \frac{1}{|I_g|}\sum_{c \in I_g} e_c, \qquad \sigma_g = \sqrt{\frac{1}{|I_g|}\sum_{c \in I_g} (e_c - \mu_g)^2},
\end{equation}

(element-wise, population standard deviation), concatenated as $[\mu_\text{VOLT}, \sigma_\text{VOLT}, \mu_\text{POWR}, \sigma_\text{POWR}, \mu_\text{VARS}, \sigma_\text{VARS}] \in \mathbb{R}^{4608}$. This is invariant to the (electrically arbitrary) number and ordering of channels within a group and gives a $>15\times$ reduction in width relative to the raw concatenation.

\subsection{Sketch Fingerprint Construction}
\label{app:sketch_fingerprints}
For each of the six 768-dimensional blocks $x_b$ from Appendix~\ref{app:type_group_pooling}, we augment the group summary with a compact fingerprint designed to recover local/transient structure that mean/std pooling smooths away. Using $\tilde{x}_b$ for the sample-wise, RMS-standardized block,

\begin{equation}\label{eq:rms_std}
\tilde{x}_b = \frac{x_b - \text{mean}(x_b)}{\sqrt{\text{mean}\big((x_b - \text{mean}(x_b))^2\big) + \varepsilon}},
\end{equation}

we create a fingerprint with two parts:
\begin{enumerate}
    \item \textit{Random sketch:} a deterministic $768 \times d$ Rademacher projection $P_b$ (entries $\pm 1/\sqrt{768}$, with signs drawn from a SHA-256 hash of the block name and $d$, so $P_b$ is reproducible without being stored), giving $\text{sketch}_b = \tilde{x}_b^\top P_b \in \mathbb{R}^d$. The ablation in \autoref{tab:psml5_rawish_sketch_appendix} sweeps $d \in \{2,4,8,16,96,160\}$. For budgets 96 and 160 the fingerprint set additionally incorporates the raw-channel sketch (the base+raw family of Section~\ref{sec:methdology:heads}); smaller budgets use the base-sketch family only.
    \item \textit{Top-$k$ energy:} a single scalar $\text{energy}_b = \log\!\big(1 + \text{mean}(\text{top-}k(|\tilde{x}_b|)^2)\big)$ summarizing the magnitude of the $k$ largest-magnitude standardized components ($k=8$ by default), capturing how peaked the block's standardized profile is.
\end{enumerate}

The fingerprint $[\text{sketch}_b,\ \text{energy}_b] \in \mathbb{R}^{d+1}$ is combined with a constant bias term, two norm-based statistics ($\log(1+\lVert x\rVert_2)$ and $\log(1+\lVert x_b\rVert_2)$), and the block's share of total squared norm, into a $(d+5)$-dimensional auxiliary vector $a_b$. Both $x_b$ and $a_b$ are then L2-renormalized and combined at a fixed energy split $\alpha$ (default $0.1$):

\begin{equation}\label{eq:energy_split}
x_b' = \frac{x_b}{\|x_b\|_2}\sqrt{1-\alpha}, \qquad a_b' = \frac{a_b}{\|a_b\|_2}\sqrt{\alpha}
\end{equation}

and the block output is $[x_b', a_b'] \in \mathbb{R}^{773+d}$. Concatenating all six blocks gives a total feature width of $6(773+d)$. The energy split $\alpha$ is fixed independent of $d$, so growing the sketch dimension trades off which auxiliary sub-components (norm statistics vs. sketch directions) share the fixed $\alpha$ energy budget, rather than growing the auxiliary vector's weight relative to the base block.

\section{Hyperparameter Search Spaces and Training Protocols}
\label{app:hparam_search}

\subsection{Classical head}
\label{app:hparam_classical}

The list of the hyperparameters of the classical head and their range can be found in \autoref{tab:hparam_classical}.

\begin{table}[htbp]
\centering
\caption{Classical head search space.}
\label{tab:hparam_classical}
\begin{tabular}{@{}ll@{}}
\toprule
Hyperparameter & Range / values \\
\midrule
Learning rate & log-uniform, $[10^{-5},\,2\times10^{-3}]$ \\
Dropout & $[0.25,\,0.5]$ \\
Weight decay & log-uniform, $[10^{-4},\,10^{-2}]$ \\
Early-stopping patience & integer, $[5,\,10]$ \\
LR decay factor & $[0.95,\,0.99]$ \\
Batch size & $\{8,16,32,64,128\}$ \\
LR schedule & $\{\text{cosine, exponential, step, none}\}$ \\
Hidden-layer architecture & $\{[64],[128],[256],[128,64],$ \\
 & $\phantom{\{}[256,128],[512,256],$ \\
 & $\phantom{\{}[128,64,32],[256,128,64]\}$ \\
\bottomrule
\end{tabular}
\end{table}

\subsection{Quantum head, core only}
\label{app:hparam_quantum_core}

The list of the hyperparameters of the core PQC for our quantum head and their range can be found in \autoref{tab:hparam_quantum_core}.

\begin{table}[htbp]
\centering
\caption{Quantum head (core only) search space.}
\label{tab:hparam_quantum_core}
\begin{tabular}{@{}ll@{}}
\toprule
Hyperparameter & Range / values \\
\midrule
Learning rate & log-uniform, $[0.5,\,3]\times 10^{-3}$ \\
Weight decay & log-uniform, $[0.5,\,5]\times10^{-3}$ \\
Dropout (classical pre-encoder) & $[0.4,\,0.7]$ \\
Pre-encoder width & integer, $[16,\,128]$ \\
Qubit count (beyond a 10-qubit baseline) & $\{-6,-4,-2,0,2,4,6\}$ \\
Circuit depth (data-reuploading layers) & $\{1,2,3,4\}$ \\
Re-uploads per layer & $\{1,2,3\}$ \\
Number of measured qubits & $\{1,2,3,4\}$ \\
Entangling-gate connectivity & $\{1,2\}$ \\
Number of parallel pre-encoders & $\{1,2\}$ \\
Post-selection & $\{\text{none, active}\}$ \\
Final linear layer & $\{\text{present, absent}\}$ \\
\bottomrule
\end{tabular}
\end{table}

\subsection{Manual refinement rounds}
\label{app:hparam_refinement}

After the initial automated search above, the quantum-head search space was revised across several successive manual rounds, in places moving outside the initial automated ranges; the final operating configuration is the one used throughout Section~\ref{sec:classification_results} and reported in the appendices.

\subsection{Engineered-feature configurations without wings}
\label{app:hparam_engineered_no_wings}

The classical head with sketch fingerprints is tuned over the same search space presented in Appendix~\ref{app:hparam_classical}, with additional fields selecting the feature-pooling and feature-transform settings for the longer, sketch-augmented input. The quantum head, core only, with sketch fingerprints does not re-search: the fixed core configuration from Appendices~\ref{app:hparam_quantum_core} and \ref{app:hparam_refinement} is reused unchanged, and only the sketch-fingerprint dimension is varied as the reported ablation axis.

\subsection{Wing training protocol}
\label{app:hparam_wings}
At each qubit count, a constant (unscheduled) learning rate is selected via a bracket search on a single seed, with the search range re-centered on each scale's own apparent optimum; the resulting core-plus-wing system is then trained with AdamW at batch size 16 for up to 500 epochs, with early-stopping patience 100 on validation balanced accuracy, repeated across four seeds under five-fold cross-validation. Data re-uploading uses $R=2$ repetitions throughout, and circuit depth is fixed per module, not growing as wings are added. All circuits are simulated noiselessly, with analytic expectation values rather than finite-shot sampling. The wing payload was upgraded partway through this process, from raw per-channel features to the raw-channel sketch described in Section~\ref{sec:methdology:heads}; both payload variants were evaluated at 16 qubits before the upgraded version was adopted for the one- and two-wing, full-scale results reported in Section~\ref{sec:qubit_scaling}.

\section{Qubit-Scaling Ladder: Full Per-Seed Results}
\label{app:qubit_ladder_full}

\autoref{tab:qubit_ladder_full} reports the individual per-seed
fold-mean balanced accuracy underlying \autoref{fig:qubit_ladder}, together with two distinct measures of spread. The $n{=}4$ column reports the standard deviation \emph{across} the four seed fold-means, indicating how reproducible the reported mean is under a re-run, since each seed re-draws both the cross-validation split and the network initialization. This is the error bar used in \autoref{fig:qubit_ladder}. The pooled column instead reports the within-seed pooled standard deviation (the root mean square of the four per-seed fold standard deviations): the typical spread of a single run's five fold scores. We report both because individual folds swing by several points (pooled $\approx 0.03$--$0.05$) while the seed-level error bars are much smaller ($\approx 0.01$); reporting only the latter would understate the run-to-run variability in any single fold's score. The pooled standard deviation is similar in magnitude across all four configurations in \autoref{tab:qubit_ladder_full}, which we take as evidence that adding wings does not destabilize evaluation, i.e.\ it does not introduce additional fold-to-fold variance beyond what the zero-wing and reference configurations already show.

\begin{table*}[!htbp]
\centering
\caption{Per-seed fold-mean balanced accuracy for the qubit-scaling
ladder and the 12-qubit reference. * marks the peak learning rate,
selected on the seed-42 curve.}
\label{tab:qubit_ladder_full}
\begin{tabular}{@{}llccccc c c@{}}
\toprule
Rung & LR & s42 & s43 & s44 & s45 & Mean$\pm$sd ($n{=}4$) & Pooled sd \\
\midrule
13q & .007 & .8296 & .8506 & .8257 & .8255 & $0.8328\pm0.0120$ & $0.0391$ \\
    & .008* & .8351 & .8534 & .8328 & .8241 & $0.8363\pm0.0123$ & $0.0448$ \\
    & .009 & .8283 & .8407 & .8283 & .8291 & $0.8316\pm0.0061$ & $0.0445$ \\
\midrule
16q & .005 & .8499 & .8558 & .8236 & .8216 & $0.8377\pm0.0177$ & $0.0500$ \\
    & .006* & .8674 & .8533 & .8365 & .8276 & $0.8462\pm0.0177$ & $0.0335$ \\
    & .007 & .8525 & .8526 & .8269 & .8273 & $0.8398\pm0.0147$ & $0.0398$ \\
    & .008\textsuperscript{aux} & .8482 & .8438 & .8342 & .8371 & $0.8408\pm0.0063$ & $0.0395$ \\
\midrule
19q & .008 & .8531 & .8730 & .8337 & .8349 & $0.8487\pm0.0185$ & $0.0478$ \\
    & .009* & .8632 & .8591 & .8513 & .8355 & $0.8523\pm0.0122$ & $0.0435$ \\
    & .010 & .8499 & .8563 & .8324 & .8299 & $0.8421\pm0.0130$ & $0.0481$ \\
\midrule
12q ref.\textsuperscript{ext} & .008 & .8254 & .8491 & .8275 & .8184 & $0.8301\pm0.0132$ & $0.0389$ \\
    & .009* & .8566 & .8317 & .8286 & .8188 & $0.8339\pm0.0161$ & $0.0341$ \\
    & .010 & .8470 & .8477 & .8166 & .8300 & $0.8353\pm0.0149$ & $0.0376$ \\
\bottomrule
\end{tabular}
\begin{flushleft}
\footnotesize\textsuperscript{aux} Common operating point (lr=.008)
shared across all rungs for the selection-free comparison in
Section~\ref{sec:qubit_scaling}; lies outside the reported search window for
16q and is not itself a candidate peak.
\textsuperscript{ext} External reference, different apparatus (no accept
qubit, no post-selection); not a ladder rung.
\end{flushleft}
\end{table*}

\section{PSML-5 Time-Series Benchmark and Sketch Fingerprint Ablation}
\label{sec:psml5_rawish_appendix}

In this appendix, we provide as reference the full values in \autoref{tab:psml5_rawish_sketch_appendix} that comprise \autoref{fig:sketch_ablation} in the main text.

\begin{table*}[!htbp]
\centering
\scriptsize
\caption{PSML-5 balanced accuracy benchmark and 12-qubit sPQC sketch-fingerprint ablation. Each row reports the best configuration per budget, five-fold mean balanced accuracy $\pm$ fold standard deviation; E3–E7 are k=48 uniform sweeps, and E2 carries a connectivity-mixing caveat. The published PSML benchmark is from Zheng et al.~\cite{zheng2022psml}.}
\label{tab:psml5_rawish_sketch_appendix}
\resizebox{\textwidth}{!}{
\begin{tabular}{llccccc}
\toprule
ID
& Representation / Setting
& Sketch Fingerprints
& Published Benchmark
& sPQC Quantum Head
& Classical MLP Head
& Q--MLP Gap \\
\midrule
B0
& PSML published benchmark~\cite{zheng2022psml}
& --
& $74.20 \pm 2.90$
& --
& --
& -- \\

E0
& Chronos Base, cross-channel mean/std fix
& --
& --
& $73.34 \pm 2.90$
& $74.43 \pm 2.20$
& $-1.09$ \\

E1
& Chronos Base, type-group mean/std
& 0
& --
& $81.93 \pm 5.00$
& $82.10 \pm 2.70$
& $-0.17$ \\

E2
& Type-group mean/std + sketch fingerprints
& 2
& --
& $82.72 \pm 3.69$
& $81.49 \pm 2.63$
& $+1.23$ \\

E3
& Type-group mean/std + sketch fingerprints
& 4
& --
& $83.90 \pm 2.04$
& $82.50 \pm 2.21$
& $+1.40$ \\

E4
& Type-group mean/std + sketch fingerprints
& 8
& --
& $84.21 \pm 3.93$
& $82.24 \pm 4.60$
& $\mathbf{+1.97}$ \\

E5
& Type-group mean/std + sketch fingerprints
& 16
& --
& $84.29 \pm 3.63$
& $82.43 \pm 2.32$
& $+1.86$ \\

E6
& Type-group mean/std + sketch fingerprints
& 96
& --
& $\mathbf{84.39 \pm 3.40}$
& $\mathbf{82.66 \pm 3.74}$
& $+1.73$ \\

E7
& Type-group mean/std + sketch fingerprints
& 160
& --
& $84.37 \pm 2.24$
& $82.41 \pm 3.54$
& $+1.96$ \\
\bottomrule
\end{tabular}
}
\end{table*}

\section{Preregistered One-Shot Test Evaluation}
\label{app:test_eval}
All results in the main text are five-fold cross-validation means on the
439-sample training split. As preregistered, after the ladder freeze we
ran a single one-shot evaluation on PSML's untouched 110-sample test
split: inference only, on the frozen fold checkpoints, with no training
or re-selection. \autoref{tab:test_eval} reports, for each rung, the
mean over the 20 per-checkpoint scores and the five-fold-ensemble score.
Test scores sit $4$--$5$ percentage points below the cross-validation
means, a generalization gap of similar size at every rung. With 110 test
samples, binomial noise is roughly $\pm3$--$4$ points, so we draw no
conclusions about rung ordering from this table; the scaling claims rest
on the cross-validation results. Relative to the published PSML-5
baseline (74.2, held-out test), the bootstrap 95\% confidence intervals
of the ensemble scores (from a $B{=}10{,}000$ sample-level bootstrap)
lie entirely above the baseline for the 13- and 19-qubit rungs; the
16-qubit lower bound (0.7435) is marginal, and the 12-qubit reference's
lower bound (0.7388) falls below it.

\begin{table}[t]
\centering
\caption{One-shot test evaluation (PSML 110-sample split), inference on
frozen checkpoints. Per-checkpoint: mean $\pm$ sd over the 20 fold
checkpoints (4 seeds $\times$ 5 folds; preregistered primary reading);
ensemble: mean $\pm$ sd over the four per-seed five-fold ensembles
(secondary). The CV column repeats the cross-validation means for
reference; Gap is per-checkpoint minus CV.}
\label{tab:test_eval}
\begin{tabular}{@{}lcccc@{}}
\toprule
Rung & Per-ckpt & Ensemble & CV & Gap \\
\midrule
12q ref. & $0.7874 \pm 0.0429$ & $0.8007 \pm 0.0150$ & $0.8339$ & $-0.0465$ \\
13q & $0.7949 \pm 0.0379$ & $0.8303 \pm 0.0179$ & $0.8363$ & $-0.0414$ \\
16q & $0.8040 \pm 0.0290$ & $0.8111 \pm 0.0157$ & $0.8462$ & $-0.0422$ \\
19q & $0.8018 \pm 0.0303$ & $0.8159 \pm 0.0152$ & $0.8523$ & $-0.0505$ \\
\bottomrule
\end{tabular}
\end{table}

\end{document}